\documentclass[preprint,aps]{revtex4}

\newcommand {\bc}{\begin{center}}
\newcommand {\ec}{\end{center}}
\newcommand {\bea}{\begin{eqnarray}}
\newcommand {\eea}{\end{eqnarray}}
\newcommand {\be}{\begin{equation}}
\newcommand {\ee}{\end{equation}}

\def\lsim{\mathrel{\rlap{\lower4pt\hbox{\hskip1pt$\sim$}}
    \raise1pt\hbox{$<$}}}               
\def\gsim{\mathrel{\rlap{\lower4pt\hbox{\hskip1pt$\sim$}}
    \raise1pt\hbox{$>$}}}                
\usepackage{graphicx}
\usepackage{color}

\begin{document}


\title{Dissipative fluid dynamics for the dilute Fermi gas at unitarity:
Anisotropic fluid dynamics}

\author{M. Bluhm and T.~Sch\"afer}

\affiliation{Department of Physics, North Carolina State University,
Raleigh, NC 27695}

\begin{abstract}
We consider the time evolution of a dilute atomic Fermi gas after
release from a trapping potential. A common difficulty with 
using fluid dynamics to study the expansion of the gas is that 
the theory is not applicable in the dilute corona, and that a naive 
treatment of the entire cloud using fluid dynamics leads to 
unphysical results. We propose to remedy this problem by
including certain non-hydrodynamic degrees of freedom, in 
particular anisotropic components of the pressure tensor, 
in the theoretical description. We show that, using this 
method, it is possible to describe the crossover from fluid
dynamics to ballistic expansion locally. We illustrate the use
of anisotropic fluid dynamics by studying the expansion of the 
dilute Fermi gas at unitarity using different functional forms
of the shear viscosity, including a shear viscosity which is 
solely a function of temperature, $\eta\sim (mT)^{3/2}$, as 
predicted by kinetic theory in the dilute limit. 

\end{abstract}

\maketitle

\section{Introduction}
\label{sec_intro}

 Considerable effort has been devoted to extracting transport
properties, in particular the shear viscosity and the spin diffusion
constant, of dilute atomic Fermi gases~\cite{Kinast:2005,Schafer:2007pr,Turlapov:2007,Cao:2010wa,Sommer:2011,Bruun:2011b,Koschorreck:2013,Elliott:2013b,Joseph:2014}.
The interest in these experiments is driven by the observation that 
strongly correlated Fermi gases can serve as model systems for
other quantum many body systems, such as high $T_c$ superconductors
or the quark-gluon plasma~\cite{Guo:2010,Schafer:2009dj,Adams:2012th}.
There are, however, two difficulties that have prevented truly 
model independent measurements of transport coefficients in trapped
systems so far. The first difficulty is that the diffusion constants 
for momentum or spin depend on the local density while the associated 
experimental observables are global measures such as the mean square 
cloud size or the total spin current. This implies the need to unfold 
the experimental data in order to obtain the density and temperature
dependence of the transport coefficients. The analogous deconvolution
problem for equilibrium quantities has been overcome using a number of 
techniques~\cite{Ho:2010,Ku:2011}, but the first study attempting to 
determine the local shear viscosity only appeared recently~\cite{Joseph:2014}. 

 The second, more serious, difficulty is that the diffusion approximation 
breaks down in the dilute part of the cloud. This problem cannot be ignored, 
because a naive application of the Navier-Stokes or the diffusion equation 
to the dilute corona leads to paradoxical behavior. Consider, for example, 
a scale invariant Fermi gas expanding after release from a harmonic 
trap~\cite{Schaefer:2009px}. In the case of a vanishing shear viscosity 
the expansion dynamics is described by an exact scaling solution of the 
Euler equation. This solution corresponds to a Hubble-like flow, in which 
the fluid velocity $\vec{u}$ is always linearly proportional to the 
distance from the center of the trap, and the temperature is only a 
function of time. We can now study how this picture is modified in 
the presence of a small dissipative term. The viscous contribution 
to the stress tensor, 
\be
\delta\Pi_{ij} = -\eta\left( \nabla_iu_j+ \nabla_j u_i-\frac{2}{3}
  \delta_{ij} \vec{\nabla}\cdot \vec{u}\right) \, ,
\ee
is a constant in space that multiplies the local shear viscosity $\eta$. 
At unitarity scale invariance implies that $\eta=n f(n/T^{2/3})$, where 
$n$ is the density, $T$ is the temperature, and $f(x)$ is a universal 
function. In the dilute limit the shear viscosity is only a function of 
temperature and not of density, $\eta={\it const}\cdot (mT)^{3/2}$. Kinetic 
theory predicts ${\it const} =15/(32\sqrt{\pi})$~\cite{Bruun:2005,Bruun:2006}. 
Since the temperature is spatially constant one concludes that $\delta
\Pi_{ij}$ goes to a constant in the dilute part of the cloud. This implies 
that there is no dissipative force, but a constant amount of dissipative 
heating per unit volume, where the energy for this infinite amount of 
heating is supplied by a heat current that flows in from spatial infinity. 

 This description is, of course, completely wrong. The mean free path in
the dilute corona is much larger than the inter-particle spacing, and
there are no collisions that could establish dissipative forces or viscous 
heating. Particles in the dilute corona are ballistically streaming. In 
order to describe the situation correctly, we have to combine a fluid 
dynamical description of the core with a weakly collisional theory of 
the corona. In this work we suggest that an efficient method for achieving 
this goal is to include certain non-hydrodynamic degrees of freedom, an 
anisotropic pressure tensor, in the theoretical description. In 
Section~\ref{sec_cross} we motivate this method by studying certain exact solutions
of the Boltzmann equation. In Section~\ref{sec_hydro} we review the 
derivation of standard, isotropic, fluid dynamics from kinetic theory, 
and in Sections~\ref{sec_ahydro} and~\ref{sec_lag} we extend this method 
to anisotropic fluid dynamics. A similar method was proposed as an 
extension of fluid dynamics to describe the early stage of a heavy-ion 
collision, see~\cite{Florkowski:2010,Martinez:2010sc}. In 
Sections~\ref{sec_num} and~\ref{sec_res} we describe numerical methods and show 
results from an anisotropic fluid dynamics code. This code is a 
generalization of the Navier-Stokes code described in~\cite{Schafer:2010dv}. 
We show that our method describes the crossover from fluid dynamics to 
free streaming both globally, for a shear viscosity of the form $\eta
\sim n$, and locally, for a shear viscosity of the form $\eta\sim 
(mT)^{3/2}$. We end with an outlook in Section~\ref{sec_out}.

\section{Global crossover from fluid dynamics to ballistic expansion}
\label{sec_cross}

 The global crossover from fluid dynamics to free streaming in the 
expansion of a trapped Fermi gas after release from the trap was studied 
in~\cite{Dusling:2011dq}, based on a set of scaling solutions to the 
Boltzmann equation obtained in \cite{Menotti:2002,Pedri:2003}.
We will use these solutions to motivate an extension of the 
fluid dynamic equations that accommodates the transition from fluid
dynamics to free streaming locally. This approach is described 
in Sect.~\ref{sec_ahydro}.

 The scaling solutions introduced in \cite{Menotti:2002,Pedri:2003}
solve the Boltzmann equation in a harmonic confinement potential and
using the Bhatnagar-Gross-Krook (BGK) approximation. This approximation 
is based on a collision term of the form $C[f]=-\delta f/\tau$, where 
$\delta f$ is the deviation of the distribution function $f$ from the 
local equilibrium distribution, and $\tau$ is the relaxation time. 
The authors of \cite{Dusling:2011dq,Menotti:2002,Pedri:2003} further 
assumed that $\tau$ is only a function of the local temperature, but 
not of the local density. In the case of two-dimensional traps
extensions of the scaling ansatz to anharmonic confining potentials 
were studied in \cite{Chiacchiera:2013}.

 Fluid dynamics corresponds to the limit $\tau\to 0$, and free streaming 
is realized as $\tau\to\infty$. In both limits the Boltzmann equation 
is solved by a distribution function of the form~\cite{Menotti:2002}
\be
\label{be_scale}
f(\vec{x},\vec{v},t) = 
  \Gamma(t) \,f_0(\vec{R}(t),\vec{U}(t)) \, ,
\ee
where 
\be 
\label{f_initial}
f_0(\vec{R},\vec{U})  \sim \exp\left(-\frac{m}{2T}
  \sum_i\left[\omega_i^2 R_i^2 + U_i^2\right]\right) 
\ee
is the initial distribution function in a harmonic potential with 
frequencies $\omega_i$ and 
\be 
\label{scale_factor}
\Gamma(t) = \prod_i\frac{1}{b_i(t)\,\theta_i(t)^{1/2}} \, , \hspace{0.3cm}
R_i(t) = \frac{x_i}{b_i(t)} \, , \hspace{0.3cm}
U_i(t) = \frac{v_i-u_i}{\theta_i(t)^{1/2}} \, , \hspace{0.3cm}
\alpha_i(t) = \frac{\dot{b}_i(t)}{b_i(t)} \, ,
\ee
with $u_i = \alpha_i(t)\,x_i$ for fixed $i$. We note that the ansatz 
in Eq.~(\ref{be_scale}) preserves the shape of the initial Boltzmann 
distribution in the cartesian directions $i=x,y,z$, and that $\theta_i$ 
plays the role of an anisotropic scale factor for the temperature. In 
the free streaming limit the solution of the Boltzmann equation is 
\be 
\label{free_stream}
\theta_i(t) = \frac{1}{b_i(t)^2} \, , \hspace{0.5cm}
b_i(t) = \left(1+\omega_i^2t^2\right)^{1/2} \, . 
\ee
In the limit of ideal fluid dynamics we find $\theta_i(t)=\bar\theta(t)$ 
with $\bar\theta(t)=[\prod_i b_i(t)]^{-2/3}$, which implies that the 
temperature is isotropic. The scale parameters $b_i(t)$ are determined by 
\be 
\label{euler_sol}
\ddot b_i(t) = \frac{\omega_i^2}{\left[\prod_j b_j(t) \right]^{2/3} b_i(t)} \, .
\ee
This equation can be solved analytically in the limit of late times and 
a strongly deformed trap, $\omega_z=\lambda\omega_\perp$ and $\omega_x=
\omega_y=\omega_\perp$ with trap deformation $\lambda\ll 1$. In this case 
one finds $b_\perp(t)\simeq \sqrt{3/2}\,\omega_\perp t$.

\begin{figure}[t]
\bc\includegraphics[width=7.5cm]{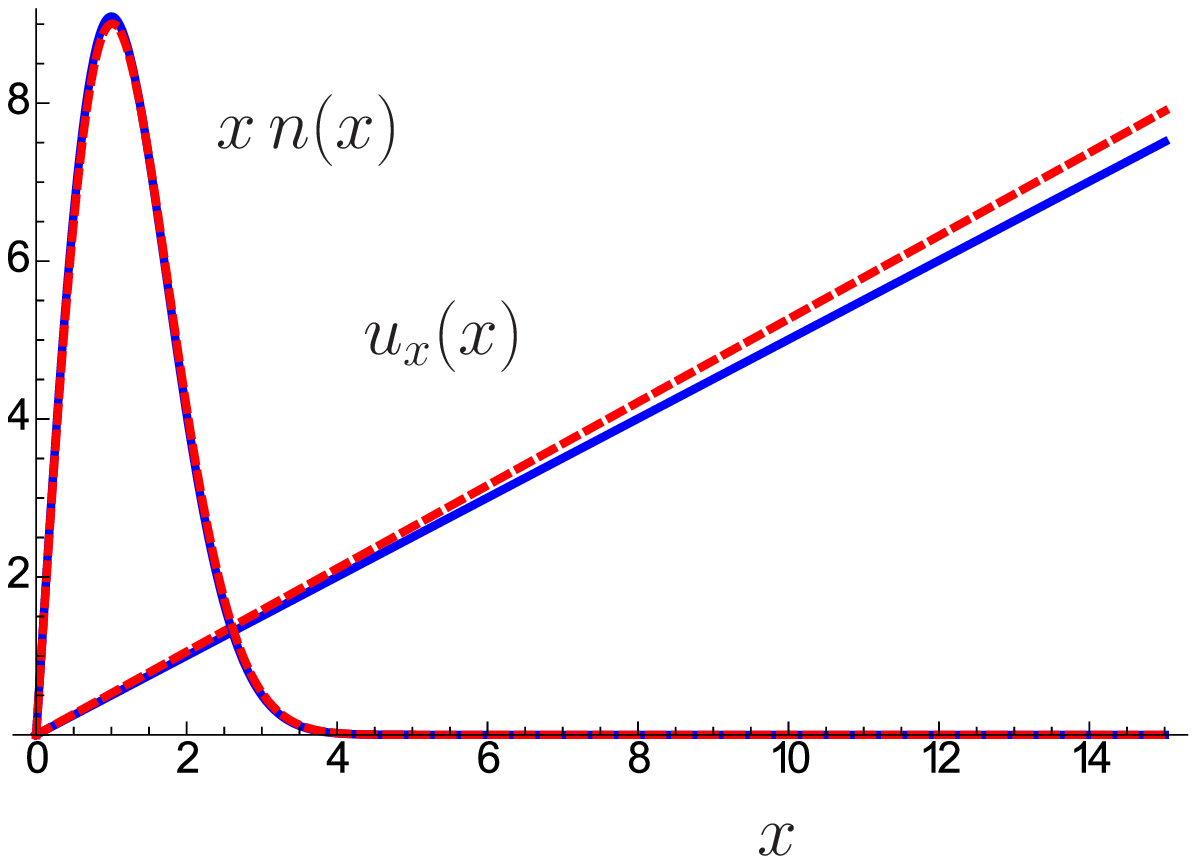}
\hspace{0.5cm}
\includegraphics[width=7.5cm]{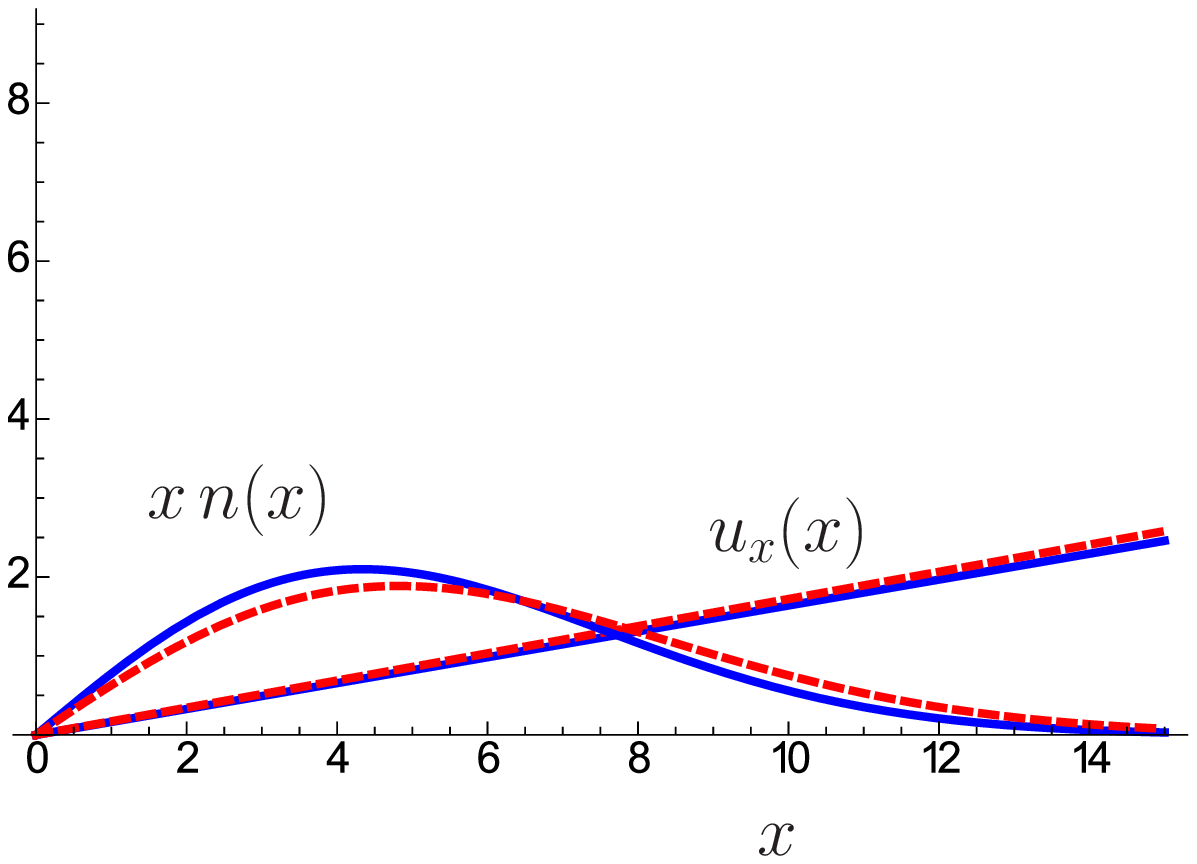}
\ec
\caption{\label{fig_exp}
Comparison between solutions of the Boltzmann equation in the limits of 
free streaming (blue solid curves) and ideal fluid dynamics (red dashed 
curves). In the left panel we show the velocity field component $u_x$ and 
the transverse density profile $xn(x)$ at an early time $t=\omega_\perp^{-1}$. 
In the right panel we show these observables at a later time $t = 6\,
\omega_\perp^{-1}$. The solutions correspond to a trap deformation $\lambda
=0.045$. The density $n$ and the velocity $u_x$ are given in arbitrary 
units, but the scales in the left and the right panel are identical.}   
\end{figure}

Solutions for the transverse flow velocity and the density distribution
in the transverse plane are shown in Fig.~\ref{fig_exp}. We observe
that the free streaming and fluid dynamic solutions are qualitatively 
similar. Transverse pressure in fluid dynamics leads to acceleration, 
which is reflected in the larger expansion velocity of the fluid dynamics 
solution in the left panel. Over time, the larger velocity shifts
the peak of the density distribution to larger radii, as shown in the 
right panel. It is interesting to note that the velocity field at late 
times is the same in free streaming and ideal fluid dynamics. The mean
velocity, that is the velocity weighted by the density, is larger in 
fluid dynamics because the maximum of the density is shifted to larger 
radii. In the limit $\lambda\ll 1$ this difference in the mean velocity 
can be understood in terms of energy conservation. In free streaming 
the internal energy of the fluid is transferred equally to kinetic 
energy in all three directions. In fluid dynamics most of the energy 
is transferred to transverse motion, and the mean velocity is larger 
by a factor $\sqrt{3/2}\simeq 1.22$.

  Figure~\ref{fig_exp} shows the difference between ideal fluid dynamics 
and free streaming in the idealized situation that the relaxation time 
is not a function of density, so that the entire cloud is either in the 
ballistic or the fluid dynamical regime. In reality a transition between 
the two regimes occurs in the dilute part of the cloud, and the transition 
region may shift during the evolution. In Sect.~\ref{sec_ahydro} we will 
study a theoretical approach that can dynamically, as a function of time 
and the spatial coordinates, accommodate the crossover from fluid dynamical 
to ballistic behavior. 

\section{Fluid dynamics from kinetic theory}
\label{sec_hydro}
 
 Before we introduce anisotropic fluid dynamics we review the derivation
of standard fluid dynamics from kinetic theory. We can view fluid dynamics 
as an effective description of a fluid that arises from the Boltzmann 
equation in the limit of a short mean free path. Consider the Boltzmann 
equation 
\be
\label{be}
\left( \partial_t + \vec{v}\cdot\vec{\nabla}_x 
                  - \vec{F}\cdot\vec{\nabla}_p \right) 
  f_p(\vec{x},t) = C[f_p]\, , 
\ee
where $f_p$ is the single-particle distribution function, $C[f_p]$ is
the collision term, $\vec{v}=\vec{\nabla}_pE_p$ the velocity of a 
particle with energy $E_p$, and $\vec{F}=-\vec{\nabla}_xE_p$ is a 
force. Using the properties of the collision term, in particular the 
conservation of particle number, energy and momentum, we can derive 
conservation laws for the conserved currents. Taking moments of the 
Boltzmann equation we find (the repeated index $j$ is summed over)
\bea
\label{hydro1}
\frac{\partial \rho}{\partial t} 
   + \vec{\nabla}\cdot \vec{\pi}  & = & 0 \, , \nonumber \\
\label{hydro2}
\frac{\partial \pi_i}{\partial t} 
   + \nabla_j\Pi_{ij} & = & 0 \, , \nonumber \\
\label{hydro3}
\frac{\partial {\cal E}}{\partial t} 
   + \vec{\nabla}\cdot\vec{\jmath}^{\,{\cal E}} & = & 0 \, . 
\eea 
The conserved charges, the mass density $\rho=mn$, the momentum
density $\vec{\pi}$, and the energy density ${\cal E}$, are given by 
\bea
\label{rho}
\rho(\vec{x},t) & = & \int d\Gamma_p\, m f_p(\vec{x},t) \, , \nonumber \\
\label{pi}
\vec{\pi}(\vec{x},t) & = & \int d\Gamma_p\, m\vec{v} f_p(\vec{x},t) 
   \, ,\nonumber \\
\label{e_dens}
{\cal E}(\vec{x},t) & = & \int d\Gamma_p\, E_p f_p(\vec{x},t) \, , 
\eea
where $d\Gamma_p=d^3p/(2\pi)^3$. The momentum density $\vec{\pi}$ is 
also the conserved current associated with the conservation of mass. 
The remaining conserved currents are the stress tensor $\Pi_{ij}$ and
the energy current $\vec{\jmath}^{\,{\cal E}}$,
\bea
\label{pi_ij_kin}
\Pi_{ij}(\vec{x},t) & = & 
\int d\Gamma_p\, p_i v_j f_p(\vec{x},t) \, , \\
\label{energycurrent_kin}
\vec{\jmath}^{\,{\cal E}} (\vec{x},t) & = & \int d\Gamma_p\, E_p 
 \left(\vec{\nabla}_p E_p\right) f_p(\vec{x},t) \, .
\eea
In order for Eqs.~(\ref{hydro3}) and~(\ref{e_dens}) to close we have 
to supply constitutive equations, that is explicit expressions for the 
conserved currents in terms of the fluid dynamical variables $\rho$, 
$\vec{\pi}$ and ${\cal E}$. In kinetic theory constitutive equations 
can be derived by expanding the distribution function around the local 
thermodynamic equilibrium distribution $f_p^0$,
\be 
 f_p = f_p^0 + \delta f_p^1 + \delta f_p^2 + \ldots \, , 
\ee
where 
\be 
\label{explform}
 f_p^0 = \exp\left(\left[\mu-E(|\vec{v}-\vec{u}|)\right]/T\right) \, ,
\ee
and $\delta f_p^n$ are terms that contain $n$'th order gradients of the 
fluid dynamical variables. The equilibrium distribution function is 
expressed in terms of intensive quantities, the local chemical potential 
$\mu(\vec{x},t)$, the temperature $T(\vec{x},t)$, and the fluid velocity 
$\vec{u}(\vec{x},t)$. From Eq.~(\ref{explform}) we can compute the 
conserved currents at zeroth order in the gradient expansion. We get 
$\vec{\pi}=\rho\vec{u}$ and 
\be 
\label{pi_ij_0}
 \Pi_{ij} = \rho u_i u_j + P\delta_{ij} \, ,
\ee
as well as $\vec{\jmath}^{\,{\cal E}} =\vec{u}\left(w+\frac{1}{2}\rho\vec{u}^2
\right)$. Here, $P$ is the pressure and $w={\cal E}^0+P$ is the enthalpy 
density, where ${\cal E}^0$ denotes the energy density in the local 
rest frame of the fluid, ${\cal E}^0={\cal E}-\frac{1}{2}\rho\vec{u}^{2}$. 
The conservation laws combined with Eq.~(\ref{pi_ij_0}) lead to the Euler 
equations of ideal fluid dynamics. The final ingredient needed to 
complete the description is an equation of state, $P=P({\cal E}_0,\rho)$. 
Using the dispersion relation of a free particle, $E(\vec{v})=\frac{1}{2}
m\vec{v}^{\,2}$, we obtain $P=\frac{2}{3}{\cal E}_0$, which agrees
with the exact result for a scale invariant fluid. 

 The local equilibrium distribution function is a solution of the 
Boltzmann equation at leading order in the Knudsen number ${\it Kn}=
l_{\it mfp}/L$, where $l_{\it mfp}$ is the mean free path and $L$ is 
the characteristic distance over which the conserved charges vary. 
At next order a solution can be found most easily by using a very 
simple form of the collision term. Using the BGK collision term
\be 
 C[f_p] = -\frac{f_p-f_p^0}{\tau} \, , 
\ee
and, again, taking the dispersion relation to be that of a free particle, 
we find 
\be 
\delta f^1_p = -\frac{m\tau f_p^0}{2T}\left(
 c_ic_j\sigma_{ij} + \left[ \frac{5T}{m}-c^2\right]
      c_k q_{k}\right) \, , 
\ee
where we have defined $\vec{c}=\vec{v}-\vec{u}$, and repeated indices
$i,j,k$ are summed over. We have also introduced the strain tensor
\be 
 \sigma_{ij} = \nabla_i u_j +\nabla_j u_i 
           -\frac{2}{3}\delta_{ij}\vec{\nabla}\cdot\vec{u} \, ,
\ee
as well as $\vec{q}=-\vec{\nabla}\log(T)$. The corresponding
corrections to the conserved currents are
\be 
\label{del_pi}
 \delta\Pi_{ij} = -\eta\sigma_{ij} \, , \hspace{1cm} 
 \delta\jmath_i^{\,{\cal E}} = u_j\delta\Pi_{ij} - \kappa \nabla_i T\, , 
\ee
where we have defined the shear viscosity $\eta=\tau P$ and the thermal 
conductivity $\kappa=\frac{5}{2}\tau P$. Incorporating the gradient 
corrections in Eq.~(\ref{del_pi}) into the conservation laws leads to 
the Navier-Stokes equation. Note that within the approximations used
here the shear viscosity and the thermal conductivity are proportional 
to one another, and the bulk viscosity $\zeta$ is zero. In general, 
$\eta$, $\zeta$ and $\kappa$ are independent parameters, but in 
a scale invariant fluid $\zeta=0$ is an exact result.

\section{Anisotropic fluid dynamics from kinetic theory}
\label{sec_ahydro}

 The gradient expansion fails in the dilute regime of the cloud. An 
obvious solution to this problem is to consider the full Boltzmann 
equation, see~\cite{Pantel:2014jfa}. Such an approach is motivated by
the observation that even though the classical Boltzmann equation is 
only justified in the dilute regime, it reproduces the fluid dynamical 
limit in the dense regime. This implies that coarse grained observables 
extracted from the Boltzmann equation are in fact more reliable than 
the kinetic theory which is used to derive them. There are, however, 
some difficulties with this approach. First of all, the Boltzmann 
equation involves a six-dimensional phase space distribution function,
and is considerably more difficult to solve than the Navier-Stokes
equation. Second, the transport properties are now encoded in a non-linear 
collision integral, which is difficult to compute from first principles, 
and not easy to parameterize in a way that allows for a shear viscosity 
which is a general function of density and temperature. And finally, it 
is difficult to incorporate the empirical equation of state. 

 An alternative approach is to use a set of fluid dynamical equations 
which is equivalent to the approach presented in the previous section
at some fixed order in the gradient expansion, but also contains 
extra, non-hydrodynamic, degrees of freedom that ensure a smooth 
crossover to the ballistic regime. Consider
\be 
 f_p = f_p^{\it an} + \delta f_p^{\prime\,1} + \delta f_p^{\prime\,2} 
   + \ldots \, , 
\ee
where 
\be 
\label{f_an}
 f_p^{\it an} = \exp\left(\frac{\mu}{T_{\it le}} 
 - \sum_a \frac{mc_a^2}{2T_a} \right) \, , \hspace{1cm}
 T_{\it le} = \left(\prod_a T_a\right)^{1/3} \, . 
\ee
The form of the anisotropic distribution function $f^{\it an}_p$ is motivated 
by the observation that, for suitable choices of $\mu$, $T_a$ and $u_a=v_a-
c_a$ Eq.~(\ref{f_an}) is an exact solution of the Boltzmann equation describing
the expansion from a harmonic trapping potential in the free streaming, 
collisionless limit, see Section~\ref{sec_cross}. In order to derive the 
conservation laws 
and constitutive equations we will use the free dispersion relation $E_p=
p^2/(2m)$. This is sufficient in order to recover the ballistic and fluid 
dynamical limits, but restricts the form of the equation of state to $P=nT
=\frac{n}{3}\sum_aT_a$. This is not a problem in the scale invariant limit, 
because the evolution equations are only sensitive to the relation $P(
{\cal E}^0)=\frac{2}{3}{\cal E}^0$, which is fixed by scale invariance.  
The full equation of state, $P=P(n,T)$, is needed to determine the initial
density profile from the equation of hydrostatic equilibrium, $\vec{\nabla}
P=-n\vec{\nabla} V$, where $V$ is the confining potential. 

 We note that the ansatz for $f^{\it an}_p$ breaks rotational invariance.
This particular ansatz is intended for analyzing the expansion of a
gas cloud from a harmonic confinement potential, where the symmetry 
axes of the potential are aligned with the cartesian coordinate system
used. Rotational symmetry can be restored by using the more general
ansatz
\be 
\label{f_an_rot}
 f_p^{\it an} = \exp\left(\frac{\mu}{T_{\it le}} 
 - \sum_{a,b} \frac{m}{2}c_a\theta_{ab}c_b \right) \, , \hspace{1cm}
 T_{\it le} = \left(\det\left(\theta^{-1}\right)\right)^{1/3} \, . 
\ee
For our purposes we will continue to use the simpler ansatz given in 
Eq.~(\ref{f_an}).

 We can use Eqs.~(\ref{rho})~-~(\ref{energycurrent_kin}) to determine 
the constitutive equations. We find $\vec{\pi}=\rho\vec{u}$ and 
\be 
\label{eps_an}
 {\cal E} = \frac{1}{2}\rho \vec{u}^2 + {\cal E}^0 \, , \hspace{0.5cm} 
 {\cal E}^0 = \frac{3}{2}P \, . 
\ee
The stress tensor is given by 
\be
\label{pi_ij_an}
 \Pi_{ij} = \rho u_i u_j + P\delta_{ij} + \delta \Pi_{ij} \, , \hspace{0.5cm} 
 \delta \Pi_{ij} = \delta_{ia}\delta_{ja}\Delta P_a \, , 
\ee
where $\Delta P_a=P_a-P$. We use the convention that repeated vector 
indices $i,j,k$ are summed over, but repeated anisotropic indices
$a,b$ are not, unless an explicit summation symbol occurs. The components 
of the energy current are 
\be
\label{j_eps_an}
 \jmath_i^{\cal E} = u_i\left(\frac{1}{2}\rho \vec{u}^2
                  + w\right) + \delta \jmath_i^{\cal E} \, , 
 \hspace{0.5cm} 
 \delta \jmath_i^{\cal E} = u_j\delta\Pi_{ij} 
              = \delta_{ia}u_a\Delta P_a \, ,
\ee
where $w={\cal E}^0+P$. In kinetic theory we also find $P=nT$ and $P_a=nT_a$. 
Combining the constitutive equations~(\ref{eps_an})~-~(\ref{j_eps_an}) with 
the conservation laws Eqs.~(\ref{hydro3}) and~(\ref{e_dens}) gives five 
equations for seven fluid dynamical variables, $\mu$, $P_{a}$ and $u_{i}$. 
We can get two additional equations by considering further moments of 
the Boltzmann equation. The conservation laws arise from taking moments 
with respect to the conserved quantities $1$, $m\vec{v}$, and $m\vec{v}^2/2$. 
Taking moments with $mv_a^2/2$ (no sum over $a$) gives
\be 
\label{mom_v_ii}
 \frac{\partial {\cal E}_a}{\partial t} 
 + \vec{\nabla}\cdot \vec{\jmath}^{\,{\cal E}}_a 
 = - \frac{\Delta P_a}{2\tau} \, , 
\ee
where we have defined 
\bea 
\label{eps_i}
 {\cal E}_a & = & \frac{1}{2}\rho u_a^2 + {\cal E}^0_a \, ,  \\
\label{j_eps_i}
 (\jmath^{\cal E}_a)_i & = & u_i\left( \frac{1}{2}\rho u_a^2 
 + {\cal E}^0_a + \delta_{ia} P\right) 
 + (\delta \jmath^{\cal E}_a)_i \, ,
\eea
and ${\cal E}^0_a=\frac{1}{2}P_a$ as well as
\be 
 (\delta \jmath^{\cal E}_a)_i = \delta_{ia}u_j\delta\Pi_{ij}
 = \delta_{ia}u_a\Delta P_a \, . 
\ee
Note that Eq.~(\ref{mom_v_ii}), when summed over $a$, gives 
the equation of energy conservation. The remaining two equations
determine the non-equilibrium pressure components $P_a$. Also note that 
we can derive additional equations based on the off-diagonal moments 
with $mv_av_b/2$ ($a\neq b$). These equations determine the off-diagonal
components of the temperature in Eq.~(\ref{f_an_rot}). 

 We will show in Section~\ref{sec_lag} that Eq.~(\ref{pi_ij_an})
reduces to the Navier-Stokes stress tensor in the limit $\tau\to 0$.
This implies that $f_p^{\it an}$ already contains all terms of order 
${\cal O}(\nabla_i u_j)$ in $\delta f_p^1$, and that $\delta f_p^{\prime\,1}$ 
only includes terms associated with heat conduction, $\delta f_p^{\prime\,1}
={\cal O}(\nabla_i T)$. It is straightforward to include these effects, but 
in the context of an expanding gas cloud heat conduction is a very small 
effect, because the initial state is isothermal and this property is 
preserved by the evolution in ideal fluid dynamics. This implies that 
gradients of the temperature are proportional to $\tau$, and, thus,
heat flow is a second order effect in the relaxation time, 
$\kappa\nabla_i T={\cal O}(\tau^2)$. 

\section{Fluid dynamical equations in Lagrangian form}
\label{sec_lag}

 In practice we solve the equations of fluid dynamics in Lagrangian
form. We introduce the comoving time derivative $D_0=\partial_0 + 
\vec{u}\cdot\vec{\nabla}$. The continuity equation can be written as 
\be
\label{rho_lag}
 D_0\rho = -\rho \vec{\nabla}\cdot \vec{u}
\ee
and the equations of momentum and energy conservation are
\bea
\label{u_lag}
 D_0 u_i & = & - \frac{1}{\rho} \left( \nabla_i P 
 + \nabla_j \delta \Pi_{ij} \right) \, , \\
\label{e_lag}
 D_0 \epsilon & = & - \frac{1}{\rho} \nabla_i \left( u_i P 
 + \delta \jmath^{\cal E}_i \right) \, , 
\eea
where we have defined the energy per mass $\epsilon={\cal E}/\rho$
and $\delta \jmath^{\cal E}_i = u_j\delta \Pi_{ij}$ as well as
$\delta\Pi_{ij}=\delta_{ia}\delta_{ja}\Delta P_a$. The equation for 
the anisotropic energy density can be written as 
\be 
\label{e_a_lag}
 D_0 \epsilon_a = - \frac{1}{\rho}
 \nabla_i \left[ \delta_{ia} u_i P + (\delta \jmath^{\cal E}_a)_i \right] 
 - \frac{1}{2\tau\rho}\Delta P_a \, , 
\ee
where $\epsilon_a={\cal E}_a/\rho$ and $(\delta \jmath^{\cal E}_a)_i 
= \delta_{ia} u_j \delta \Pi_{ij}$. In standard fluid dynamics we view 
$\rho$, $u_i$ and ${\cal E}$ as the fluid dynamical variables. Their time 
evolution is governed by Eqs.~(\ref{rho_lag})~-~(\ref{e_lag}), and in 
order to determine the RHS of Eqs.~(\ref{u_lag}) and~(\ref{e_lag}) we use 
the equation of state $P({\cal E}^0)$ with ${\cal E}^0={\cal E}-\frac{1}{2}
\rho\vec{u}^2$, where in a scale invariant fluid $P({\cal E}^0)=\frac{2}{3}
{\cal E}^0$. In anisotropic fluid dynamics we have two extra variables, 
${\cal E}_1$ and ${\cal E}_2$ with $\sum_a {\cal E}_a={\cal E}$. Their 
time evolution is governed by Eq.~(\ref{e_a_lag}), and $P_a$ is given 
by the anisotropic equation of state $P_a({\cal E}^0_a)=2{\cal E}^0_a$ 
with ${\cal E}^0_a={\cal E}_a-\frac{1}{2}\rho u_a^2$. Note that $P=
\frac{1}{3}\sum_a P_a$ satisfies the isotropic equation of state.

 In the previous section we argued that in the limit $\tau\to 0$ anisotropic 
fluid dynamics reduces to Navier-Stokes viscous fluid dynamics. We expect,
in particular, that the dissipative correction to the stress tensor $\delta 
\Pi_{ij}=\delta_{ia}\delta_{ja}\Delta P_a$ approaches $\delta\Pi_{ij}= -\eta 
\sigma_{ij}$ (for $i=j$) with $\eta=\tau P$. To see this we rewrite 
Eq.~(\ref{e_a_lag}) as 
\be 
\label{e_a_lag_re}
 \Delta P_a = - 2\tau\rho \left(D_0 \epsilon_a + \frac{1}{\rho}
 \nabla_i \left[ \delta_{ia} u_i P + (\delta \jmath^{\cal E}_a)_i\right]\right)
\ee
and solve for $\Delta P_a$ at leading order in $\tau$. This implies that 
in evaluating  ${\cal E}^0_a$ and $(\delta \jmath^{\cal E}_a)_i$ we can 
replace $P_a$ by $P$, so that $\epsilon_a=\frac{1}{3}\epsilon-\frac{1}{6}
\vec{u}^2+\frac{1}{2} u_a^2$ and $(\delta \jmath^{\cal E}_a)_i=0$. We use
the equations of ideal fluid dynamics to compute $D_0 \epsilon$ and 
$D_0 u_i$ and find 
\be
\label{proof}
 \Delta P_a = \tau P \left(\frac{2}{3}\vec{\nabla}\cdot \vec{u} - 
           2 \,\nabla_a u_a\right) + {\cal O}(\tau^2)
 = -\tau P \sigma_{aa} +  {\cal O}(\tau^2) \, .
\ee
This result shows that anisotropic fluid dynamics relaxes to the 
Navier-Stokes equation with 
\be 
\eta = \tau P\, ,
\ee
where $\tau$ is the relaxation time.
Note that the expression in Eq.~(\ref{pi_ij_an}) does not reproduce the
off-diagonal components of $\delta\Pi_{ij}$ in Navier-Stokes theory. In 
order to study flows in which these terms are non-zero we have to start 
from the more general ansatz in Eq.~(\ref{f_an_rot}) and consider moments 
of the Boltzmann equation with $mv_av_b/2$ for $a\neq b$.

\section{Anisotropic fluid dynamics: Numerical method and choice of units}
\label{sec_num}

 We have implemented anisotropic fluid dynamics as an extension of the 
Navier-Stokes code described in~\cite{Schafer:2010dv}. The Navier-Stokes
code solves the advection equations in Lagrangian coordinates. A Lagrangian 
time step is followed by a piecewise parabolic remap onto an Eulerian 
grid. The algorithm is based on the PPMLR (Piecewise-Parabolic Method, 
Lagrangian-Remap) scheme developed by Colella and Woodward~\cite{Colella:1984} 
and implemented as a multi-dimensional method in the VH1 code written 
by Blondin and Lufkin~\cite{Blondin:1993}. The main modification is that 
we add the fluid dynamical variables $P_a$ and ${\cal E}_a$ and solve 
Eq.~(\ref{e_a_lag}). We solve for all three components of $P_a$ and 
verify that $\frac{1}{3}\sum_a P_a$ agrees with $P$. Since Eq.~(\ref{e_a_lag}) 
is a relaxation equation for $\Delta P_a$ we have to choose the time step as 
\be 
 \Delta t = \min_x \left(c_s\frac{\Delta x}{2},u_i\frac{\Delta x}{2},
    \tau\right) \, , 
\ee
where $c_s$ is the local speed of sound, $\Delta x$ is the grid spacing, 
and $\tau=\eta/P$ is the local relaxation time. The first two constraints
arise from the condition that disturbances emerging from opposite faces 
of a fluid cell cannot interact during a time step. The third constraint 
ensures that the relaxation time equation is stable. The condition $\Delta 
t\leq \tau$ implies that the simulation becomes inefficient if $\tau$ is 
very small, which is close to the limit of ideal fluid dynamics. In 
principle this can be addressed by using the analytic result given 
in Eq.~(\ref{proof}), but we have not done so in the present work. 

 We have studied the evolution of a unitary Fermi gas after release from a 
harmonic trap. The trapping potential is $V(x)=\frac{1}{2}m\omega_i^2x_i^2$ 
with $\omega_x=\omega_y=\omega_\perp$ and $\omega_z=\lambda\omega_\perp$. 
We use dimensionless variables for distance, time and velocity based on 
the following system of units~\cite{Schafer:2010dv}
\be
\label{x_0_def}
 x_0 = (3N\lambda)^{1/6}\left(\frac{2}{3m\omega_\perp}\right)^{1/2} \, , 
 \hspace{0.5cm}
 t_0 = \omega_\perp^{-1} \, ,
 \hspace{0.5cm}
 u_0 = x_0\omega_\perp \, . 
\ee
The unit of density is $n_0=x_0^{-3}$. The corresponding units for energy 
density, pressure, and temperature are given by 
\be
\label{P_0_def}
 {\cal E}_0 = \frac{m\omega_\perp^2}{x_0} \, ,
 \hspace{0.5cm}
 P_0 = \frac{m\omega_\perp^2}{x_0} \, ,
 \hspace{0.5cm}
 T_0 = m\omega_\perp^2x_0^2 \, . 
\ee
Finally, the unit of the shear viscosity is 
\be 
\label{eta_0_def}
 \eta_0 = \frac{m\omega_\perp}{x_0} \, . 
\ee
In the high temperature limit the initial density is a Gaussian.
The central density is given by 
\be 
\label{n_0}
 n(0) = n_0 \frac{N\lambda}{\pi^{3/2}}\left(\frac{E_F}{E_0}\right)^{3/2} \, ,
\ee
where $N$ is the number of particles, $E_F=(3N\lambda)^{1/3}\omega_\perp$
is the Fermi energy, and $E_0$ is the total energy per particle of the 
trapped gas. Moreover, it is convenient to normalize the dimensionless 
central density $n(0)/n_0$ to one. This means we also divide the density
by the dimensionless factor $(N\lambda)/\pi^{3/2}\cdot (E_F/E_0)^{3/2}$
in equ.~(\ref{n_0}). The normalized dimensionless shear viscosity is 
\be 
 \bar\eta = \frac{\eta}{\eta_0}\frac{n_0}{n(0)} \, . 
\ee
In the following we will consider a shear viscosity of the form 
$\eta=\alpha_n n+\alpha_T(mT)^{3/2}$. The corresponding dimensionless 
shear viscosity is $\bar\eta=\bar\alpha_n\bar{n}+\bar\alpha_T\bar{T}^{3/2}$ 
with 
\be
\label{alpha_bar}
 \bar\alpha_n = \frac{3}{2}\frac{\alpha_n}{(3N\lambda)^{1/3}} \, , 
 \hspace{0.5cm}
 \bar\alpha_T = \frac{4\pi^{3/2}}{3}\frac{\alpha_T}{(3N\lambda)^{1/3}}
 \left(\frac{E_0}{E_F}\right)^{3/2} \, .
\ee
Kinetic theory predicts that in the high temperature limit $\alpha_n=0$ 
and $\alpha_T=15/(32\sqrt{\pi})$. In the anisotropic fluid dynamics
framework the shear viscosity is determined by the relaxation time. 
The dimensionless relaxation time is 
\be 
\bar{\tau}= \frac{\tau}{t_0} = \frac{\bar\eta}{\bar{P}}\, . 
\ee
Equation (\ref{alpha_bar}) shows that, in dimensionless units, the 
number of particles $N$ only appears together with the viscosity 
coefficient. This implies that the ideal evolution is independent
of $N$, and that for a given viscosity dissipative effects are
smaller for a larger number of particles.

\begin{figure}[t]
\bc\includegraphics[width=9.5cm]{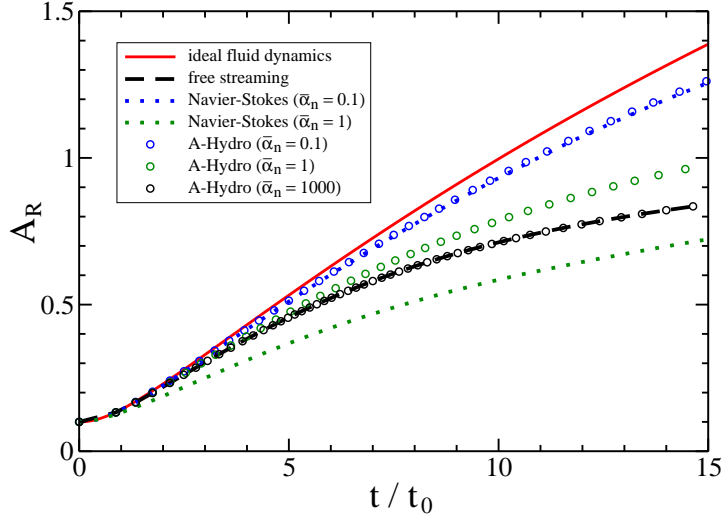}\ec
\caption{\label{fig_A_R}
This figure shows the evolution of the aspect ratio $A_R$ as a function
of time $t$ in dimensionless units, as explained in the text. The initial
trap deformation is $A_R(0)=\lambda=0.1$ and the initial energy is $E_0/E_F=1$. 
The red solid curve shows the evolution in ideal fluid dynamics, and the 
black dashed curve is the free streaming limit. The remaining curves were 
obtained using viscous fluid dynamics with different values of $\bar\alpha_n$ 
for the shear viscosity $\bar\eta=\bar\alpha_n\bar{n}$. The dotted curves 
show Navier-Stokes results for $\bar\alpha_n=0.1,\,1$ and the data points 
are the corresponding predictions from anisotropic fluid dynamics. In the 
case of anisotropic fluid dynamics we also show the result for $\bar\alpha_n
=1000$, which is close to the free streaming limit. }
\end{figure}

\section{Anisotropic fluid dynamics: Results}
\label{sec_res}

 We first consider the case $\alpha_T=0$ and study the dependence on 
$\alpha_n$. Figure~\ref{fig_A_R} shows the time evolution of the aspect 
ratio $A_R(t)=\langle r_\perp^2\rangle/\langle r_z^2\rangle$, defined by 
the ratio of mean squared transverse and longitudinal cloud radii, for 
different values of the shear viscosity, $\bar\alpha_n=0.1,\,1,\,1000$. 
For comparison we also show the result in ideal fluid dynamics, the free 
streaming limit, and the solution of the Navier-Stokes equation for 
$\bar\alpha_n=0.1,\,1$. We consider a Gaussian initial condition, which 
corresponds to a solution of the hydrostatic equation in the case of 
an equation of state of a free gas, $P=nT$. At $t=0$ the aspect ratio is 
given by the trap deformation, $A_R(0)=\lambda$. Pressure gradients 
preferentially accelerate the fluid in the transverse direction, and 
$A_R(t)$ grows as a function of time. Viscous effects counteract the 
expansion in the transverse direction, and accelerate the fluid in 
the longitudinal direction, reducing the value of $A_R(t)$. 

\begin{figure}[t]
\bc\includegraphics[width=9.5cm]{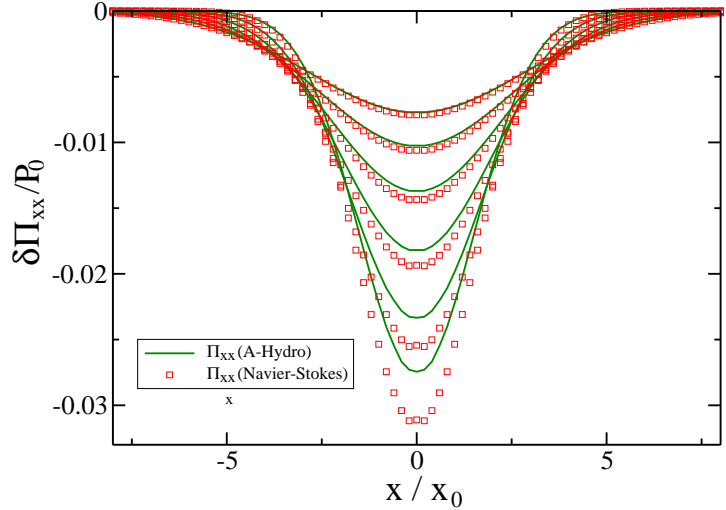}\ec
\caption{\label{fig_dpi_an}
This figure shows the $xx$ component of the dissipative correction to
the stress tensor $\delta\Pi_{xx}(x,0,0)$ in Navier-Stokes theory (green 
solid curve) and in anisotropic fluid dynamics (red squares) for different 
times $t/t_0=0.75-2.0$ in steps of $\Delta t=0.25\,t_0$. The magnitude of 
the viscous stresses decreases with time. The calculation was performed with 
a density dependent shear viscosity $\bar\eta=\bar\alpha_n\bar{n}$ and 
$\bar\alpha_n=0.15$. We used a trap deformation $\lambda=0.045$ and 
an initial energy $E_0/E_F=3$.}   
\end{figure}
 
 For the smallest value of the shear viscosity, $\bar\alpha_n=0.1$, we 
find good agreement between anisotropic fluid dynamics and Navier-Stokes 
theory, as expected from Eq.~(\ref{proof}). For larger values of $\bar
\alpha_n$ anisotropic fluid dynamics predicts that dissipative effects 
saturate, and that $A_R(t)$ approaches the free streaming limit. In 
Navier-Stokes theory, on the other hand, dissipative effects continue 
to grow with $\bar\alpha_n$ and the evolution of $A_R(t)$ becomes 
arbitrarily slow. 

 More details are provided by Fig.~\ref{fig_dpi_an}. In this figure, we 
compare the dissipative corrections to the stress tensor in Navier-Stokes 
theory and in anisotropic fluid dynamics. We focus on the $xx$ components
$\delta\Pi^{NS}_{xx}=-\eta\sigma_{xx}$ and $\delta\Pi^{AH}_{xx}=\Delta P_x$
at different times during the evolution of the expanding gas cloud. The 
calculation was performed for $\bar\alpha_n=0.15$, so that the aspect ratio 
$A_R(t)$ shows good agreement between Navier-Stokes theory and anisotropic 
fluid dynamics. Note that $\delta\Pi^{NS}_{xx}$ in Fig.~\ref{fig_dpi_an} was 
computed using the velocity field in anisotropic fluid dynamics. We observe 
that the two dissipative stress tensors are indeed very close, and that the 
agreement improves at late times. This indicates that the equations of 
anisotropic fluid dynamics contain second order terms in $\tau$ that describe 
the relaxation of the stress tensor to the Navier-Stokes 
limit~\cite{Chao:2011cy,Schaefer:2014xma}. 

\begin{figure}[t]
\bc\includegraphics[width=9.5cm]{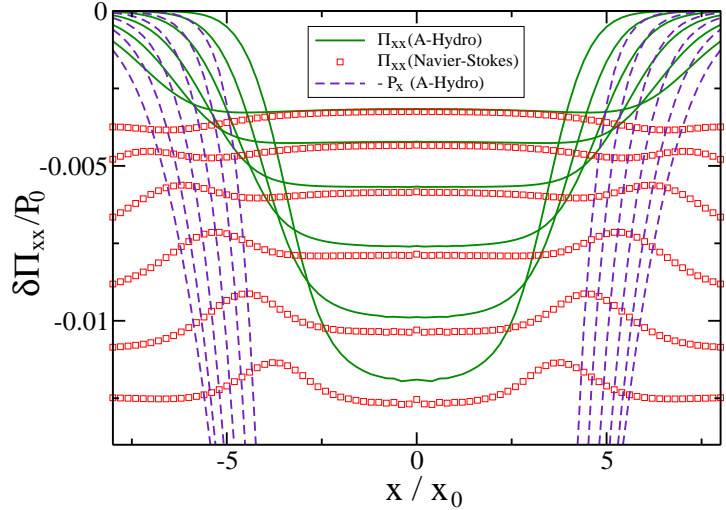}\ec
\caption{\label{fig_dpi_at}
Same as Fig.~\ref{fig_dpi_an} for a temperature dependent shear viscosity
$\bar\eta=\bar\alpha_T\bar{T}^{3/2}$ with $\bar\alpha_T=0.06$. The blue
dashed line shows the negative of the anisotropic pressure component 
$P_x$.}   
\end{figure}

 Figure~\ref{fig_dpi_at} demonstrates that anisotropic fluid dynamics
can be applied to the case of a purely temperature dependent shear
viscosity, $\eta=\alpha_T (mT)^{3/2}$, for which Navier-Stokes fluid
dynamics fails. We observe that in the center of the cloud the two
dissipative corrections to the stress tensor are close, in particular 
at late times. In the corona, however, $\delta\Pi^{NS}_{xx}$ and $\delta
\Pi^{AH}_{xx}$ are very different. As explained in Section~\ref{sec_intro} 
the dissipative contribution to the Navier-Stokes stress tensor is 
approximately constant in space. In contrast, the dissipative contribution 
to the stress tensor in anisotropic fluid dynamics goes to zero in the 
dilute part of the cloud. As we can see from the blue dashed curves in 
Fig.~\ref{fig_dpi_at} this happens in the regime where the dissipative 
stresses are comparable to the total pressure of the fluid, $|\delta
\Pi^{NS}_{xx}| \sim P_x$. This condition, corresponding to the point where
the dashed blue line intersects the red symbols, signals the breakdown
of Navier-Stokes theory. 

 We note that $-\nabla_x\delta\Pi^{AH}_{xx}$ 
corresponds to a force that points towards the center of the cloud, 
and reduces the transverse expansion. We also observe that at late 
times $\sigma_{xx}$, which measures the slope of the velocity field, 
is smaller in the center than in the corona. Viscous forces push on 
the center of the cloud and slow it down, while the ballistic corona 
is lifting off.

\begin{figure}[t]
\bc\includegraphics[width=9.5cm]{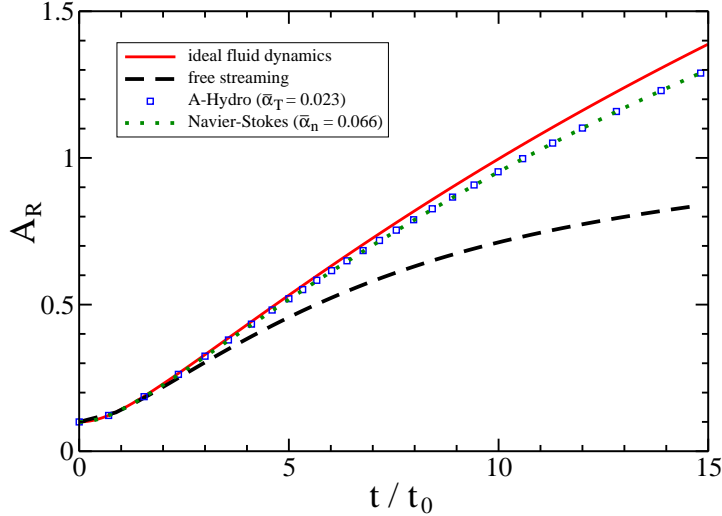}\ec
\caption{\label{fig_A_R_alt}
This figure shows the time evolution of the aspect ratio $A_R$ 
from a simulation using anisotropic fluid dynamics with $\bar\alpha_T
=0.023$ in $\bar\eta=\bar\alpha_T\bar T^{3/2}$ and $\lambda=0.1$, $E_0/E_F
=1$ (blue squares). For comparison we also show the result in ideal 
fluid dynamics (red solid curve) and the case of purely ballistic 
expansion (black dashed curve). The green dotted curve shows a fit 
to the $\bar\alpha_T=0.023$ result based on Navier-Stokes theory 
with $\bar\alpha_n = 0.066$ in $\bar\eta=\bar\alpha_n\bar n$.}
\end{figure}

 In anisotropic fluid dynamics the viscous stresses are concentrated in 
the center of the gas cloud even if $\eta\sigma_{ij}$ is not localized. We 
therefore expect that the time evolution of the aspect ratio $A_R(t)$ can be 
described by an effective density dependent shear viscosity $\eta\sim n$ even 
if the microscopic shear viscosity is only a function of temperature. This 
approximation has been used to analyze experimental data on the expansion of 
trapped Fermi gases near unitarity~\cite{Cao:2010wa,Elliott:2013b}. In 
Fig.~\ref{fig_A_R_alt} we show that for a given initial temperature, and 
for a suitably chosen value of $\alpha_n$, the evolution of $A_R(t)$ is 
indeed essentially indistinguishable between the two cases $\eta\sim n$ 
and $\eta\sim (mT)^{3/2}$. 

\begin{figure}[t]
\bc\includegraphics[width=8.5cm,clip=true]{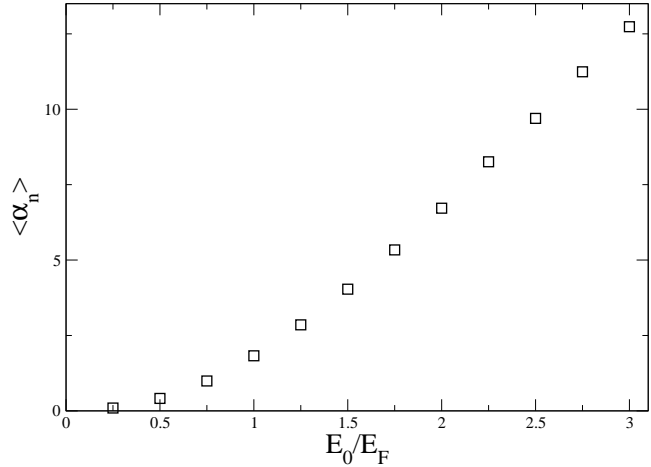}\ec
\caption{\label{fig_a_n}
This figure shows the dependence of $\langle\alpha_n\rangle$, the effective 
trap averaged ratio $\eta/n$, on the initial energy per particle in 
units of the Fermi energy $E_F$. A precise definition of $\langle\alpha_n
\rangle$ is given in the text. The calculation was performed for a gas
cloud of $2\cdot 10^5$ particles with an initial trap deformation of
$\lambda=0.045$. The microscopic shear viscosity is given by the kinetic 
theory result in the high temperature limit, $\eta=15/(32\sqrt{\pi})
(mT)^{3/2}$.}   
\end{figure}

 In order to resolve this ambiguity and determine the full microscopic 
dependence of $\eta$ on $n$ and $T$ we have to study the 
sensitivity of the effective $\alpha_n$ on the initial temperature
$T$ or the initial energy $E_0/E_F$. In the literature, the value
of $\alpha_n$ which describes the time evolution of $A_R(t)$ at a given 
initial temperature is referred to as the trap averaged value
of $\eta/n$, denoted $\langle\alpha_n\rangle$~\cite{Cao:2010wa}. In 
Fig.~\ref{fig_a_n} we show the dependence of $\langle\alpha_n\rangle$ 
on $E_0/E_F$ for the case $\eta=15/(32\sqrt{\pi})(mT)^{3/2}$. We
consider a Gaussian initial condition, so that $E_0=3T$. We observe
that the growth of $\langle\alpha_n\rangle$ is not simply proportional
to $E_0^{3/2}$. This is because the evolution is not only sensitive to
the temperature dependence of the shear viscosity, $\eta\sim T^{3/2}$, but 
also to the temperature dependence of the relaxation time, $\tau\sim
T^{1/2}$, and the temperature dependence of the effective relaxation
volume. 

 We note that as a consequence of the complicated dependence
of $\langle\alpha_n\rangle$ on the system size and lifetime the 
result is not universal, which means that $\langle\alpha_n\rangle$
depends on the number of particles $N$ and the trap deformation $\lambda$. 
In the present work we will not attempt to perform a detailed analysis
of the experimental data obtained in~\cite{Elliott:2013b,Joseph:2014}. 
This will require implementing a non-axially symmetric confining 
potential, a non-Gaussian initial density distribution, a realistic
equation of state $P(n,T)$, and a shear viscosity which is a function
of both $n$ and $T$. It is nevertheless interesting to consider a 
rough comparison between our results and the analysis in~\cite{Joseph:2014}. 
At $E_0/E_F=3.14$ we find $\langle\alpha_n\rangle = 13.49$, compared to 
$\langle\alpha_n\rangle = 19.63\pm 0.54$ reported in~\cite{Joseph:2014}.
The discrepancy is an indication of the magnitude of the effects due to 
the trap geometry, contributions beyond the dilute limit of the
equation of state and the shear viscosity, and possible shortcomings 
of our description in the regime where the size of the fluid dynamical
core shrinks to zero, and a kinetic treatment of the entire cloud is 
more appropriate. 

\section{Conclusions and outlook}
\label{sec_out}

 In this work we have shown that by including non-hydrodynamic
degrees of freedom it is possible to achieve a smooth transition
between fluid dynamics and the ballistic limit. There are a number
of interesting applications and theoretical issues that remain to be 
investigated. 

\begin{itemize}
\item Other fluid dynamic problems: The anisotropic fluid dynamics
approach, described by equ.~(\ref{rho_lag}-\ref{e_a_lag}), provides 
a general scheme for addressing problems in which the transition
regime from fluid dynamics to ballistic behavior plays a role. 
In this work we have applied this method to the expansion of a 
unitary Fermi gas from a harmonic trap, but the applicability of 
the approach is clearly much broader, including Bose or classical 
gases, as well as other experimental observables, such as collective 
modes. 

\item Restoring rotational invariance: In this work we only
considered the ``cartesian'' ansatz in Eq.~(\ref{f_an}). In order 
to restore full rotational invariance we have to start from 
Eq.~(\ref{f_an_rot}) and derive the corresponding fluid dynamical 
equations. 

\item Relation to second order fluid dynamics: In our numerical 
simulations we observed that anisotropic fluid dynamics contains some 
effects that appear at second order in the gradient expansion of fluid 
dynamics, in particular a finite viscous relaxation time. It will be 
interesting to make this more precise, and extend the equations of motion 
to complete second order accuracy. 

\item More accurate treatment of the Knudsen limit: We have shown that
anisotropic fluid dynamics reproduces the Navier-Stokes equation at 
order ${\cal O}(\tau)$, where $\tau$ is the relaxation time in the 
BGK approximation to the Boltzmann equation. In the opposite limit, 
$\tau\to\infty$, anisotropic fluid dynamics provides an exact solution 
of the Boltzmann equation for $\tau=\infty$. However, at ${\cal O}
(1/\tau)$ anisotropic fluid dynamics does not provide an exact solution 
of the Boltzmann equation, only a solution at second order in an expansion 
in moments of the distribution function with respect to momentum. The 
reliability of this approximation can be studied by comparing with the 
exact numerical solutions obtained in~\cite{Pantel:2014jfa}.

\item Extension to superfluid hydrodynamics: The experimental results
obtained in~\cite{Joseph:2014} also cover the regime $T<T_c$, where 
$T_c$ is the critical temperature for superfluidity. In order to 
extract the shear viscosity in this regime we have to extend anisotropic
fluid dynamics to the superfluid (two-fluid) regime. 

\item Extension to spin diffusion: The problem related to the dilute 
regime also affects the extraction of the spin diffusion constant. 
It will be interesting to study whether our method can be extended
to the case of charge and spin diffusion. 
\end{itemize}

 Our immediate goal is to use the method developed in this work to 
extract the local shear viscosity $\eta(n,T)=nf(n/T^{2/3})$ from the data 
presented in~\cite{Joseph:2014}. This can be achieved by inverting 
the dependence of the aspect ratio $A_R(E_0,t)$ as a function of the 
initial energy on the function $f(x)$. As explained in the previous 
section, this will 
require implementing a non-Gaussian initial density distribution as
well as considering a non-axially symmetric trap, and a realistic equation of 
state $P(n,T)$. A natural starting point for unfolding the full density
and temperature dependence of the shear viscosity is the reconstruction
presented in~\cite{Joseph:2014}. At this point we have not implemented
a superfluid version of the anisotropic fluid dynamics method, and we
are limited to the regime $T>T_c$. 
 
 Acknowledgments: This work was supported in parts by the US Department 
of Energy grant DE-FG02-03ER41260. We would like to thank James Joseph
and John Thomas for many useful discussions. We would also like to 
thank John Blondin for help with the VH1 code. 


\end{document}